\documentclass[reqno,12pt]{article}
\usepackage{amsmath,amsfonts,amssymb,amsthm,amstext,amscd,eucal,xcolor}
\usepackage[all]{xy}
\usepackage{amsmath, amssymb}
\usepackage{mathtools}
\newcommand{\mathsym}[1]{{}} 
\usepackage[colorlinks=true,
            linkcolor=blue,
            urlcolor=blue,
            citecolor=blue]{hyperref}
\usepackage{enumerate}

\usepackage{graphicx}% Include figure files
\usepackage{amsmath} \usepackage{amssymb}
\usepackage{bm}% bold math
\usepackage{cite,hyperref}
\usepackage{mathtools}
\usepackage{amsmath} \usepackage{amssymb}
\usepackage{bm}% bold math
\usepackage{graphicx}
\usepackage[active]{srcltx}
\makeatletter \@addtoreset{equation}{section}

\makeatletter\renewcommand\section{\@startsection {section}{1}{\z@}%
                                   {-3.5ex \@plus -1ex \@minus -.2ex}%nn
                                   {2.3ex \@plus.2ex}%
                                   {\normalfont\large\bfseries}}
\renewcommand\subsection{\@startsection{subsection}{2}{\z@}%
                                     {-3.25ex\@plus -1ex \@minus -.2ex}%
                                     {1.5ex \@plus .2ex}%
                                     {\normalfont\bfseries}}

\newcommand{\be}{\begin{equation}}
\newcommand{\ee}{\end{equation}}
\newcommand{\bea}{\begin{eqnarray}}
\newcommand{\eea}{\end{eqnarray}}
\newcommand{\bse}{\begin{subequations}}
\newcommand{\ese}{\end{subequations}}
\newcommand{\beqa}{\begin{eqnarray}}
\newcommand{\eeqa}{\end{eqnarray}}
\newcommand{\beqar}{\begin{eqnarray*}}
\newcommand{\eeqar}{\end{eqnarray*}}
\newcommand{\bi}{\begin{itemize}}
\newcommand{\ei}{\end{itemize}}
\newcommand{\bn}{\begin{enumerate}}
\newcommand{\en}{\end{enumerate}}
\newcommand{\ba}{\begin{array}}
\newcommand{\ea}{\end{array}}
\newcommand{\bc}{\begin{center}}
\newcommand{\ec}{\end{center}}

\definecolor{darkgreen}{rgb}{0,0.3,0}
\definecolor{darkblue}{rgb}{0,0,0.3}
\definecolor{darkred}{rgb}{0.7,0,0}
\definecolor{VioletRed4}{rgb}{0.55,0.13,0.32}
\definecolor{VioletRed}{rgb}{0.82,0.13,0.56}
\definecolor{VioletRed2}{rgb}{0.93,0.23,0.55}

\makeatother
\begin{document}

\newcommand{\email}[1]{\footnote{\href{mailto:#1}{#1}}}

\title{\bf\Large{One-loop Photon's Effective Action in the Noncommutative Scalar QED$_{3}$}}

\author{\bf{M.~Ghasemkhani}\email{ghasemkhani@ipm.ir} $^{a}$, \bf{R.~Bufalo}\email{rodrigo.bufalo@ufla.br} $^{b}$,  \bf{V.~Rahmanpour}\email{v.rahmanpour@mail.sbu.ac.ir} $^{a}$ and M.~Alipour\email{moj.alipour@yahoo.com} $^{a}$ \\\\
%\vspace{1cm}
%\normalsize
\textit{\small$^a$  Department of Physics, Shahid Beheshti University,  G.C., Evin, Tehran 19839, Iran}\\
\textit{\small $^b$   Departamento de F\'isica, Universidade Federal de Lavras,}\\
\textit{\small Caixa Postal 3037, 37200-000 Lavras, MG, Brazil}\\
}
%%%%%%%%%%%%%%%%%%%%%%%%%%%%%%%%%%%%%%%%%%%%%%%%%%%%%%%%%%%%%%%%%%%%%%%%%%%%%%%%%%%%%%%%
%\date{\today}
\maketitle

\begin{abstract}
In this paper, we consider the evaluation of the effective action for photons coupled to charged scalar fields in the framework of a $(2+1)$-dimensional noncommutative spacetime.
In order to determine the noncommutative Maxwell Lagrangian density, we follow a perturbative approach, by integrating out the charged scalar fields, to compute the respective graphs for the vev's $\left\langle AA \right\rangle$, $\left\langle AAA \right\rangle$ and $\left\langle AAAA \right\rangle$.
Surprisingly, it is shown that these contributions are planar and that, in the highly noncommutative limit, correspond to the Maxwell effective action and its higher-derivative corrections.
It is explicitly verified that the one-loop effective action is gauge invariant, as well as under discrete symmetries: parity, time reversal, and charge conjugation.
Moreover, a comparison of the main results with the noncommutative QED$_{3}$ is established.
In particular, the main difference is the absence of parity violating terms in the photon's effective action coming from integrating out the charged scalar fields.

\end{abstract}

%\vspace{0.5in}

%%%%%%%%%%%%%%%%%%%%%%%%%%%%%%%%%%%%%%%%%%%%%%%%%%%%%%%%%%%%%%%%%%%%%%%%%%%%%%%%%%%%%%%%%%
\setcounter{footnote}{0}
\renewcommand{\baselinestretch}{1.05}  %Line spacing
%%%%%%%%%%%%%%%%%%%%%%%%%%%%%%%%%%%%%%%%%%%%%%%%%%%%%%%%%%%%%%%%%%%%%%%%%%%%%%%%

%\addtocontents{toc}{\protect\setcounter{tocdepth}{2}}
\newpage
\tableofcontents
%%%%%%%%%%%%%%%%%%%%%%%%%%%%%%%%%%%%%%%%%%%%%%%%%%%%%%%%%%%%%%%%%%%%%%%%%%%%%%%%%%%%%%%%%%%
\section{Introduction}
\label{sec1}

In recent years a great amount of attention has been paid in the analysis and calculation of covariant effective action for different types of quantum fields, exploring the diversity of new interactions that mainly depend on the spin of the fields involved as well as the spacetime dimensionality \cite{Bonora:2016otz, Bonora:2017ykb, Quevillon:2018mfl}.
One may say that the canonical example of a complete analysis is the Euler-Heisenberg effective action \cite{Heisenberg:1935qt}, where quantum effects from QED are responsible to induce nonlinear interactions among photons.
Moreover, the effective action framework has served as an important tool to explore different point of views about the quantum gravitational theory, where the Einstein-Hilbert action is augmented by metric and/or torsion fields higher-order terms \cite{Buchbinder:1992rb,Masud}.

Naturally, since the framework of effective action is a powerful tool, there is a great expectation that this approach can be used to make contact with modern phenomenology of physics beyond the standard model.
The main idea behind this formulation is that at energies below some cutoff scale $\mu$, \footnote{That may signal symmetry violation, for instance Lorentz symmetry violation.} all the effects of the massive degrees of freedom above $\mu$ can be encoded as new interactions among the fields remaining active below $\mu$.
The effective action approach has been extensively used to the study of Lorentz violating field theories, where the energy scale $\mu$ is related to the Planck energy scale $E_{\rm Pl}$ (or length $\ell_{\rm Pl}$) where our notion of smooth geometry is expected to break \cite{ref53,Bluhm:2005uj}.
In this case, the current understanding is that the low energy Lorentz violating terms come as quantum corrections from heavy modes \cite{Borges:2013eda,Borges:2016uwl}.

Although the majority of studies of Lorentz violating field theories is developed in a four-dimensional spacetime, there are considerable interests in the description of
three-dimensional (3D) ones \cite{Charneski:2008hy, Nascimento:2014owa, Casana:2015hda}.
Besides the algebraic richness of odd dimensional spacetimes, one may say that the most appealing aspect of 3D field theories is the UV finiteness in some models.
This feature might provide an ambiguity free description of Lorentz violation, allowing thus a close contact of violating effects with physical planar phenomena.
In particular, it is worth recall the example of the description of quantum hall fluids in terms of noncommutative geometry \cite{Susskind:2001fb,Douglas:2001ba}.

Over the past two decades, field theories defined in a noncommutative (NC) geometry have been considered as one of the most prominent candidates presenting Lorentz violation to make contact with quantum gravity phenomenology \cite{Douglas:2001ba,AmelinoCamelia:2008qg}.
Within this description, the noncommutativity measurement parameter is related to a length scale $\ell_{\rm nc} \sim \sqrt{\theta}$.
In one hand, this length scale can be seen as a manifestation of the discreteness of the spacetime, presenting a smooth profile in the UV region \cite{Arzano:2017uuh}.
On the other hand, this same scale is responsible for introducing instabilities in the dispersion relations of the fields, the so-called UV/IR mixing \cite{Matusis:2000jf}.
%However, this type of modification in the dispersion relations can be interesting to some applications.

NC field theories have been studied through the effective action approach, where the behavior of the new couplings was deeply analyzed \cite{Vassilevich:2005vk}, where the presence of UV/IR mixing in the 1PI functions signals that applying the usual Wilsonian field theory notions and techniques to NC QFT's one should be careful.
This type of analysis was also developed to two and three-dimensional NC models
\cite{Ghasemkhani:2013bs, Ghasemkhani:2017bjp, Chu:2000bz, Banerjee:2007ua, Bufalo:2014ooa}.
These studies of effective action in 3D models were exclusive to the coupling of gauge and fermion fields, no much attention has been paid to the case involving scalar fields, in particular the case of spinless charged fields interacting with photons.

In one hand, it is of physical significance to study scalars in 3D field theories independently of fermions in condensed matter systems, as in Quantum Hall systems, since we have scalar quasiparticle excitations.
On the other hand, recently 3D versions of fermionization/bosonization have also been introduced \cite{Hsin:2016blu,Benini:2017dus}.
In these studies, it was discussed the duality between nonspin
Chern-Simons theory and a spin Chern-Simons theory, exploring precisely the spin structure of the given models.
In this sense, the present work could be the first step in extending such analysis to the NC case.
Motivated by these facts, we will analyze throughout the paper to what extent the spin of the matter fields can change the effective action when charged scalar and fermion fields are considered in the presence of the spacetime noncommutativity.
A straightforward result is that in the case of the 3D scalar quantum electrodynamics (scalar QED$_{3}$), it is not possible to generate the parity odd Chern-Simons terms, showing thus that the dynamics of the 3D gauge field is significantly different in the presence of either charged scalar or fermion fields. It is well known that the presence of the degree of freedom associated with the spin changes in most of the cases only the magnitude of physical quantities, e.g. the beta function \cite{Ghasemkhani:2016zjy} and electron's magnetic moment \cite{Panigrahi:2004cf}.

In this paper we discuss the effective action for the photon in the scalar QED in the noncommutative three-dimensional spacetime.
In Sec.~\ref{sec2} we present an overview of the scalar QED, where the charged scalar fields are minimally coupled with the photons.
There we define the main aspects regarding the Moyal product used in our analysis,\footnote{The noncommutativity we will be using in the paper is
defined by the algebra $[\hat{x}_\mu,\hat{x}_\nu] = i \theta_{\mu\nu}$. So in order to construct a noncommutative field theory, using the Weyl-Moyal (symbol)
correspondence, the ordinary product is replaced
by the Moyal star product as defined below.}
we also discuss the content of discrete symmetries in the NC 3D spacetime.
In addition, all the Feynman rules are presented for the propagators and 1PI vertices.
Section \ref{sec4} is focused in the perturbative computation of the relevant graphs corresponding to the one-loop effective action for the photon gauge field.
It is also discussed the generation of higher-derivative terms, similarly to the
Alekseev-Arbuzov-Baikov effective Lagrangian for non-Abelian fields.
In Sec.~\ref{sec5} we establish a comparison of the obtained results for the effective action in the NC-scalar QED to those of ordinary NC-QED, exploring the part played by the spin in these cases.
We present our final remarks in Sec.~\ref{conc}.

%%%%%%%%%%%%%%%%%%%%%%%%%%%%%%%%%%%%%%%%%%%%%%%%%%%%%%%%%%%%%%%%%%%%%%%%%%%%%%%%%%%%%%%%%%%%%%%%%%%%%%%%%%%%%%%%%%%%%%%%%%%%%%%%%%%%%%%%%%%%%%%%%%%%%%%%%%%%%%%%%%%%%%%%%

\section{The model}
\label{sec2}
In this section, we introduce the model and fix our notation.
The noncommutative extension of the bosonic electrodynamics is described by the following action
\begin{equation}
S=\int d^{3}x\Big[\left(D_{\mu}\phi\right)^{\dagger}\star D^{\mu}\phi-m^{2}\phi^{\dagger}\star \phi\Big],
\label{eq:a1}
\end{equation}
this functional action consists of the interaction of charged scalar fields minimally coupled with an external gauge field.
We consider the covariant derivative form in the fundamental representation $D_{\mu}\phi=\partial_{\mu}\phi+ieA_{\mu}\star\phi$.
This action is invariant under the infinitesimal gauge transformation
\begin{equation}
\delta A_{\mu}=\partial_{\mu}\lambda+ie[A_{\mu},\lambda]_{\star}~,\quad \delta\phi=ie\lambda\star\phi,
\label{eq:a2}
\end{equation}
where $[\,\, ,\,]_{\star}$ is the Moyal bracket.
Moreover, the Moyal star product between the functions $f$ and $g$ is defined as
\begin{equation}
f\left( x\right)\star g\left(x\right) = f\left( x\right) \exp \left(\frac{i}{2}\theta ^{\mu \nu}
\overleftarrow{\partial_\mu}
 \overrightarrow{\partial_\nu}\right) g\left( x\right),
 \label{eq:a3}
\end{equation}
where $\theta^{\mu\nu}=-\theta^{\nu\mu}$ are constant parameters that measure the noncommutative structure of the space-time.
In order to avoid unitarity violation, we assume that $\theta^{0i}=0$, hence we have only one nonzero independent component $\theta^{12}$ in our model.

It is worth mentioning that although the couplings \eqref{eq:a1} are simply modified by the presence of a nonplanar phase due to the Moyal product, the noncommutativity of spacetime coordinates shows its importance in the computation of the one-loop effective action for the gauge field, where nonlinear self-couplings are present solely due to the NC framework.
The one-loop effective action for the gauge field
can be readily obtained by integrating out the charged scalar fields of \eqref{eq:a1}
\begin{equation}
e^{i\Gamma_{\rm eff}[A]}=\int D\phi^{\dagger} D\phi~e^{-i\int d^{3}x~\phi^{\dagger}\star(D^{2}+m^{2})\star\phi}.
\end{equation}
Using the Gaussian functional integration formulas for the case of interacting charged scalar fields, we can write the noncommutative 1PI effective action as below
\begin{equation}
i\Gamma_{\rm eff}[A]={\rm Tr}\ln\left[\frac{(i\partial_{\mu}-eA_{\mu})\star(i\partial^{\mu}-eA^{\mu})\star-m^{2}}{-\partial^{2}-m^{2}}\right],
\end{equation}
where ${\rm Tr}$ is a sum over eigenvalues of the operator inside the bracket which can also be evaluated in momentum space.
Similarly to the description of one-loop effective action for the gauge field in the case of NC-QED \cite{Bufalo:2014ooa}, one can show that $\Gamma_{\rm eff}[A]$ has a convergent series expansion in coupling constant $e$. From a diagrammatic point of view, it includes the one-loop graphs contributing to the gauge field $n$-point functions which is considered as
\begin{equation}
\Gamma_{\rm eff}[A]={\cal{S}}_{\rm eff}[AA]+{\cal{S}}_{\rm eff}[AAA]+{\cal{S}}_{\rm eff}[AAAA]+\cdots.
\label{eq:aa}
\end{equation}
However, the functional $\Gamma_{\rm eff}[A]$, in comparison to the NC-QED case, has more graphs due to the presence of an additional interacting vertex.

Moreover, it is important to emphasize that as we will show in our model,
similarly to the case of NC-QED \cite{Ghasemkhani:2017bjp}, the one-loop effective action for the  photons is completely planar.
Explicitly, in the evaluation of the one-loop diagrams with an arbitrary number of external legs of photons, for energies below the mass scale $m$, only planar diagrams contribute. This means the absence of IR/UV mixing.

\subsection{Discrete symmetries}
\label{sec2.1}

Since we are interested in computing the one-loop effective action for the photon, it is useful to analyze the behavior of the original action \eqref{eq:a1} under discrete symmetries: parity, charge conjugation and time reversal.
This study will allow us to determine which of them may be anomalous in the obtained results for the one-loop order effective action.

\begin{itemize}
\item{\emph{Parity}}

Parity transformation in $d=2+1$ is defined as $x_{1} \rightarrow -x_{1}$ and $x_{2}\rightarrow x_{2}$, in this case we have that the field $\phi$ is even under parity, and the components of the gauge field $A_{\mu}$ behave as $A_{0}\rightarrow A_{0}$, $A_{1}\rightarrow -A_{1}$ and $A_{2}\rightarrow A_{2}$.
Moreover, we observe from the NC algebra that the $\theta$ parameter changes under this transformation as $\theta^{12}\rightarrow -\theta^{12}$ .
With these considerations, it is easy to show that the whole of the action \eqref{eq:a1} is parity invariant.
%%%%%
\item{\emph{Time Reversal}}

Under time reversal, we have that $x_{0} \rightarrow -x_{0}$. In this case, the components of the gauge field behave as $\left( A_{0}, A_i\right) \rightarrow \left( A_0,-A_i\right)$.
By demanding that the scalar field does not change $\phi \to \phi $, and that necessarily the NC parameter transforms as $\theta^{12}\rightarrow -\theta^{12}$ under time reversal, we are left with a $T$-invariant action.
%%%%
\item{\emph{Charge Conjugation}}

As we know, the behavior of the gauge field under charge conjugation is given by $A_{\mu}\rightarrow -A_{\mu}$ for any space-time dimensionality. Taking the scalar field to be unchanged under $C$, and the transformation for the NC parameter $\theta\rightarrow-\theta$, we conclude that the action \eqref{eq:a1} is $C$-invariant.
\end{itemize}

%%%%%%%%%%%%%%%%%%%%%%%%%%%%%%%%%%%%%%%%%%%%%%%
%%%%%%%%%%%%%%%%%%%%%%%%%%%%%%%%%%%%%%%%%%%%%%%

\subsection{Propagators and vertex functions}
\label{sec3}

In order to discuss the computation of the perturbative effective action, we must determine the basic propagators and 1PI vertex functions.
From the functional action described in \eqref{eq:a1}, we can obtain the bosonic propagator
\begin{equation}
{\cal{D}}(p) = \frac{i}{{p^{2}-m^{2}}},
 \label{eq:b1}
\end{equation}
the cubic vertex $\left \langle A \,\phi \, \phi^\dagger \right \rangle$
\begin{align}
\Gamma^\mu(p,q)= -ie\left(p+q\right)^{\mu}\exp\Big(\frac{i}{2} p \wedge q\Big),
 \label{eq:b2}
\end{align}
and the quartic vertex $\left \langle AA\,\phi \,\phi^\dagger \right \rangle$
\begin{align}
\Lambda^{\mu\nu}(p,q,s)=2ie^{2}\eta^{\mu\nu}\exp{\Big(\frac{i}{2}k \wedge s\Big)}\cos\Big(\frac{p \wedge  q}{2}\Big),
 \label{eq:b3}
\end{align}
where we have introduced the notation $p \wedge q = p_\mu \theta^{\mu\nu}q_\nu$.
A straightforward difference of the scalar and fermionic electrodynamics is the presence of the quartic vertex $\left \langle AA\phi \phi^\dagger \right \rangle$, which increases significantly the number of the one-loop graphs.
Moreover, the scalar vertices are rather simpler due to the absence of the Dirac $\gamma$ matrices, resulting in a much simpler algebraic analysis.

%%%%%%%%%%%%%%%%%%%%%%%%%%%%%%%%%%%%%%%%%%%%%%%%%%%%%%%%%%%%%%%%%%%%%%%
%%%%%%%%%%%%%%%%%%%%%%%%%%%%%%%%%%%%%%%%%%%%%%%%%%%%%%%%%%%%%%%%%%%%%%%

\section{Perturbative Effective Action}
 \label{sec4}

Now that we have determined the basic Feynman rules for the 1PI functions, we shall proceed to the computation of the one-loop diagrams related to the effective action for the gauge field.
For this purpose, we shall compute along this section the respective contributions: the free part of the effective action $\left\langle AA \right\rangle$, and the interacting parts for the cubic vertex $\left\langle AAA \right\rangle$ and quartic vertex $\left\langle AAAA \right\rangle$.
In general, the final results of our analysis related to the graphs contributing to $\left\langle AA \right\rangle$, $\left\langle AAA \right\rangle$ and $\left\langle AAAA \right\rangle$ vertices shall be a function of $e$, $\tilde{p}_\mu = \theta_{\mu \nu } p^\nu$ and $p^2/m^2$.

To highlight the effects of noncommutativity in the low energy effective   action and the photon two, three and four-point functions, while we take the external momenta $p$ such that $p^2/m^2\ll 1$, we also take the highly noncommutative limit, i.e., the low-energy regime $p^2/m^2 \to 0$ while $\tilde{p}$ is kept finite.
In this limit the noncommutative (planar) phase factors, which are a function of $\tilde{p}$, remain finite.
Moreover, we shall focus our attention on those terms of order $m^{-1}$.
We present by complementarity, at the next to leading order, the terms of order $m^{-3}$ that correspond to higher-derivative corrections.
%%%%%%%%%%%%%%%%%%%%%%%%% %%%%%%%%%%%%%%%%%%%%%%%%%%%%%%%%%%%%%%
 \subsection{One-loop $\left\langle AA \right\rangle$ part}
%%%%%%%%%%%%%%%%%%%%%%%%% %%%%%%%%%%%%%%%%%%%%%%%%%%%%%%%%%%%%%%

\begin{figure}[t]
\vspace{-1.2cm}
\includegraphics[height=6\baselineskip]{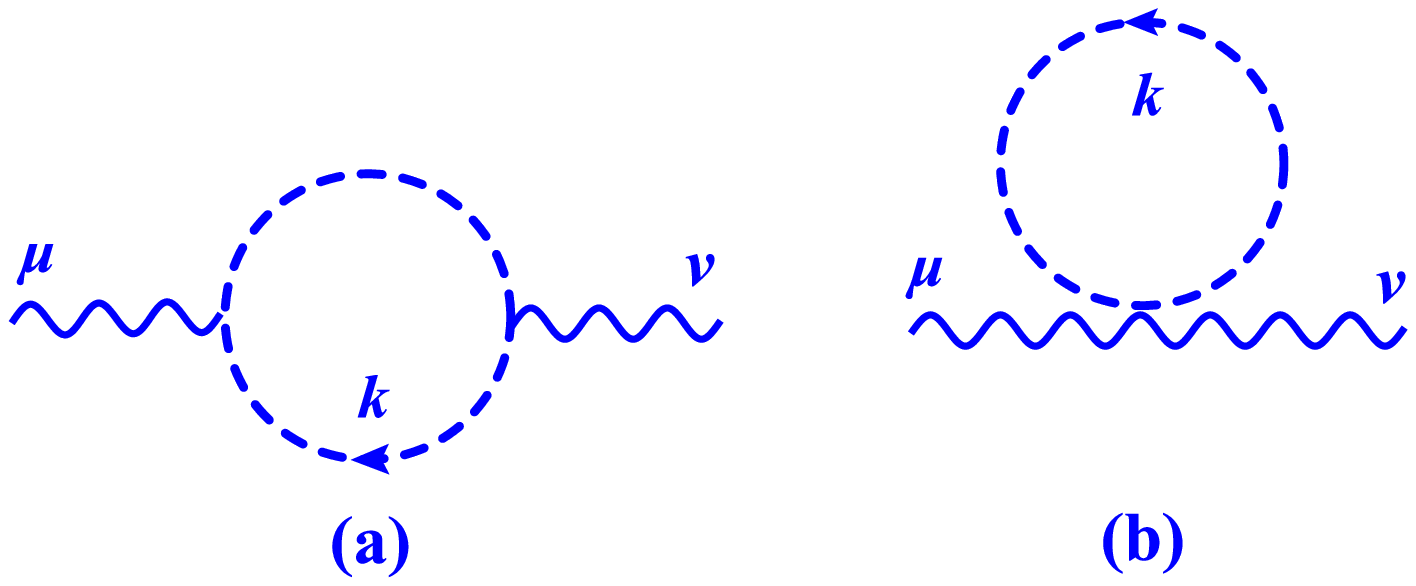}
 \centering\caption{Relevant graphs for the induced $AA$-term.}
\label{oneloop1}
\end{figure}

From the Feynman rules we can compute the one-loop contribution to the $AA$-term corresponding to the free part of the photon effective action.
The two diagrams contributing at this order are depicted in Fig.~\ref{oneloop1}, which the respective expressions have the form
\begin{align}
\Pi^{\mu\nu}_{(a)}(p)  &= e^{2} \int \frac{d^{d}k}{(2\pi)^{d}} \frac{(p+2k)^{\mu}(p+2k)^{\nu}}{[(p+k)^{2}-m^{2}][k^{2}-m^{2}]}, \nonumber \\
\Pi^{\mu\nu}_{(b)}(p) &= -e^{2} \int \frac{d^{d}k}{(2\pi)^{d}} \frac{2\eta^{\mu\nu}[(p+k)^{2}-m^{2}]}{[(p+k)^{2}-m^{2}][k^{2}-m^{2}]},
 \label{eq:c1}
\end{align}
so that the full contribution is written as
\begin{align}
\Pi^{\mu\nu}(p) &= e^{2}\int \frac{d^{d}k}{(2\pi)^{d}} \frac{(p+2k)^{\mu}(p+2k)^{\nu}-2\eta^{\mu\nu}[(p+k)^{2}-m^{2}]}{[(p+k)^{2}-m^{2}][k^{2}-m^{2}]}.
 \label{eq:c2}
\end{align}

A first comment is that this piece is completely planar, carrying no noncommutative effects.
The explicit computation is straightforward using dimensional regularization.
After some algebraic calculation, we can consider the low-energy limit, $p^2/m^2 \to 0$, resulting into
\begin{align}
\Pi^{\mu\nu}(p) =\frac{ie^{2}}{48\pi m}\left(p^{\mu}p^{\nu}-\eta^{\mu\nu}p^{2}\right).
 \label{eq:c3}
\end{align}
Moreover, for the next to leading order contribution, ${\cal{O}}(m^{-3})$, we find that
\begin{equation}
\Pi^{\mu\nu}_{\rm hd}(p)=\frac{ie^{2}}{960\pi m^{3}}\left(p^{\mu}p^{\nu}-\eta^{\mu\nu}p^{2}\right)p^{2}.
 \label{eq:c4}
\end{equation}
These two terms Eqs.~\eqref{eq:c3} and \eqref{eq:c4} satisfy straightforwardly the Ward identity, $p_\mu \Pi^{\mu\nu}=0$, as we expected.
We can determine the respective contribution to the effective action by means of
\begin{equation}
i{\cal{S}}_{\rm eff}[AA]=\int\int d^{3}x_{1}d^{3}x_{2}~A_{\mu}(x_{1})\Gamma^{\mu\nu}(x_{1},x_{2})A_{\nu}(x_{2}),
 \label{eq:c5}
\end{equation}
where ${\cal{S}}_{\rm eff}[AA]$ is the quadratic part of the effective action $\Gamma[A]$ in \eqref{eq:aa}. Here, we have defined by simplicity
\begin{equation}
\Gamma^{\mu\nu}(x_{1},x_{2})=\int\frac{d^{3}p}{(2\pi)^{3}} e^{-i p\cdot (x_{1}-x_{2})}\Pi^{\mu\nu}(p ).
 \label{eq:c6}
\end{equation}
%%%%%%%%%%%%%%%%%%%%%%%%%%%%%%%%%%%%
%%%%%%%%%%%%%%%%%%%%%%%%%%%%%%%%%%%%

After some algebra, the quadratic part of the induced effective action for the photon, considering \eqref{eq:c3} and \eqref{eq:c4}, is given by
\begin{align}
i{\cal{S}}_{\rm eff}[AA]&=-\frac{ie^{2}}{48\pi m}\int d^{3}x\Big(\partial_{\mu}A_{\nu}\partial^{\mu}A^{\nu}-\partial^{\mu}A_{\nu}\partial^{\nu}A_{\mu}\Big) \nonumber \\
&+\frac{ie^{2}}{960\pi m^{3}}\int d^{3}x\Big(\partial_{\mu}A_{\nu}\Box\partial^{\mu}A^{\nu}-\partial^{\mu}A_{\nu}\Box\partial^{\nu}A_{\mu}\Big).
 \label{eq:c7}
\end{align}
As we have previously mentioned, the first term of the expression \eqref{eq:c7} corresponds to the kinetic part of the noncommutative Maxwell action, ${\cal O}(m^{-1})$, while the second term is the higher-derivative correction to the kinetic term, of order ${\cal O}(m^{-3})$.
Moreover, the obtained result does not contain any noncommutativity effect, since the produced phase factors cancel for $n=2$.
It is worth noticing the absence of the parity odd Chern-Simons term in the scalar QED$_{3}$, which in turn is generated in the fermionic electrodynamics due to the algebraic structure of the two-dimensional  realization of $\gamma$ matrices.

%%%%%%%%%%%%%%%%%%%%%%%%% %%%%%%%%%%%%%%%%%%%%%%%%%%%%%%%%%%%%%%%%%%%%%%%%%%%
\subsection{One-loop $\left\langle A AA\right\rangle$ vertex}
%%%%%%%%%%%%%%%%%%%%%%%%%%%%%%%%%%%%%%%%%%%%%%%%%%%%%%%%%%%%%%%%%%%%%%%%%%%%

\begin{figure}[t]
\vspace{-1.2cm}
\includegraphics[height=8\baselineskip]{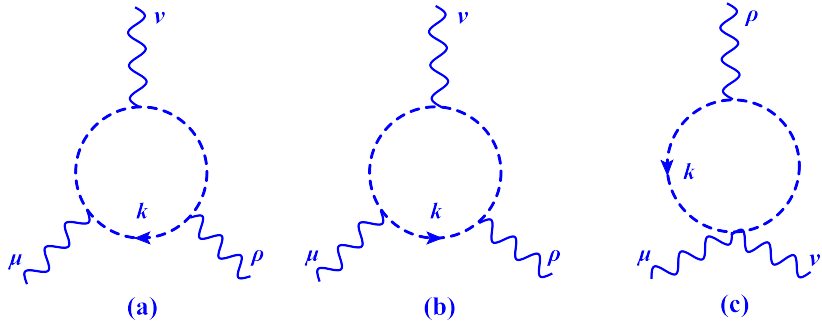}
 \centering\caption{Relevant graphs for the induced $AAA$-term.}
\label{oneloop2}
\end{figure}

The relevant graphs for the $\left\langle A AA\right\rangle$ part of the effective action are shown in Fig.~\ref{oneloop2}.
However, in order to determine correctly the full contribution to the effective action, it is necessary to consider all different permutations of the external bosonic lines of the given graphs.
It is easy to see that the diagram (a) has an additional contribution (b), corresponding to a permutation of the external photon legs, which has an equivalent structure but with a reversed momentum flow, which comes exactly from the S-matrix expansion at the order of $e^{3}$.
With help of the Feynman rules, we can easily write the relevant expression for the sum of the graphs (a) and (b)
\begin{equation}
\Pi^{\mu\nu\rho}_{(a+b)}(p,q)=2i e^{3}\int \frac{d^{d}k}{(2\pi)^{d}} \frac{(p+2k)^{\mu}(2p+2k+q)^{\nu}(p+q+2k)^{\rho}}{[(p+k)^{2}-m^{2}][(p+q+k)^{2}-m^{2}][k^{2}-m^{2}]} \sin\big(\frac{p \wedge q}{2}\big),
 \label{eq:c8}
\end{equation}
which is a planar quantity, we can see that its integrand is independent of the noncommutativity.
The contribution from the graph (c) also has a simple planar structure, which is given by the expression
\begin{equation}
\Pi^{\mu\nu\rho}_{(c)}(p,q)= -e^{3}\int \frac{d^{d}k}{(2\pi)^{d}} \frac{\eta^{\mu\nu}(p+q+2k)^{\rho}}{[(p+q+k)^{2}-m^{2}][k^{2}-m^{2}]} \cos\big(\frac{p \wedge q}{2}\big).
 \label{eq:c9}
\end{equation}

Since the graph (c) is planar, one can perform straightforward manipulations to show that this contribution is identically zero, i.e. $\Pi^{\mu\nu\rho}_{(c)}=0$, for any value of the external momenta. Hence the full contribution for the $\left\langle A AA\right\rangle$ vertex reads
\begin{align}
\Pi^{\mu\nu\rho}(p,q)=2i e^{3}\int \frac{d^{d}k}{(2\pi)^{d}} \frac{(2k+p)^{\mu}(2k+p+s)^{\nu}(s+2k)^{\rho}}{[(p+k)^{2}-m^{2}][(s+k)^{2}-m^{2}][k^{2}-m^{2}]} \sin\big(\frac{p \wedge q}{2}\big).
 \label{eq:c10}
\end{align}
%%%%%%%%%%%%%%%%%%%%%%%%%%%%%%%%%%%%%%%%%%%%%%%%%%%%%%%%%%%%%%%%%%%%%%%
%%%%%%%%%%%%%%%%%%%%%%%%%%%%%%%%%%%%%%%%%%%%%%%%%%%%%%%%%%%%%%%%%%%%%%%%%

The computation of the loop integral is lengthy but straightforward using dimensional regularization, and in the low-energy limit $p^{2},q^{2}\ll m^{2}$, we find that
\begin{equation}
\Pi^{\mu\nu\rho}(p,q)=\frac{e^{3}}{12\pi m}\bigg[\big(p-q\big)^{\rho}\eta^{\mu\nu}-\big(2p+ q\big)^{\nu}\eta^{\mu\rho} +\big(p+2 q\big)^{\mu}\eta^{\nu\rho}\bigg] \sin\big(\frac{p \wedge  q}{2}\big).
 \label{eq:c11}
\end{equation}
Here, we notice that the Eq.~\eqref{eq:c11} corresponds exactly to the standard Feynman vertex of the 3-photon interaction term in the NC spacetime.
Moreover, in the next to leading order, ${\cal{O}}(m^{-3})$, we have the contribution from the higher-derivative terms
\begin{align}
\Pi^{\mu\nu\rho}_{\rm hd}(p,q)=-\frac{e^3}{240\pi m^3}
\bigg\{&\eta^{\mu\nu}\Big[ p^2(2q-p)^{\rho}+q^2(q-2p)^{\rho}+(p.q)(q-p)^{\rho}\Big]\nonumber\\
+&\eta^{\mu\rho}\Big[p^2(4p+2q)^{\nu}+q^2(3p+q)^{\nu}+(p.q)(4p+q)^{\nu}\Big]\nonumber\\
%%%%%%
-& \eta^{\nu\rho}\Big[p^2(p+3q)^{\mu}+q^2(2p+4q)^{\mu}+(p.q)(p+4q)^{\mu}\Big]\nonumber\\
+&p^{\mu}q^{\rho}(q-p)^{\nu}+p^{\rho}q^{\mu}(q-p)^{\nu}-p^{\mu}
p^{\rho}(2p+q)^{\nu}+q^{\mu}q^{\rho}(p+2q)^{\nu}\bigg\}\sin\big(\frac{p\wedge q}{2}\big).
 \label{eq:c12}
\end{align}
This expression corresponds to the higher-derivative correction to the 3-photon vertex.

It is important to observe that in the commutative limit, the graphs (a) and (b) cancel each other, so that the induced 3-photon vertex is completely removed in the scalar QED.
We can understand this result from the charge conjugation invariance of the scalar QED in any space-time dimension, known as Furry's theorem, that forbids the presence of an odd number of photon lines in the case of commutative theory.
Another important aspect from our analysis is the absence of the Chern-Simons self-coupling $\epsilon_{\mu\nu\lambda}A^{\mu}\star A^{\nu}\star A^{\lambda}$ for the noncommutative scalar QED$_{3}$ effective action \eqref{eq:c11}, which is only generated in the case of fermionic electrodynamics \cite{Bufalo:2014ooa}.

%%%%%%%%%%%%%%%%%%%%%%%%%%%%%%%%%%%%%%%%%%%%%%%%%%%%%%%%%%%%%%%%%%
\subsection{One-loop $\left\langle AA AA \right\rangle$ vertex}
%%%%%%%%%%%%%%%%%%%%%%%%%%%%%%%%%%%%%%%%%%%%%%%%%%%%%%%%%%%%%%%%%%

\begin{figure}[t]
\vspace{-1.2cm}
\includegraphics[height=8\baselineskip]{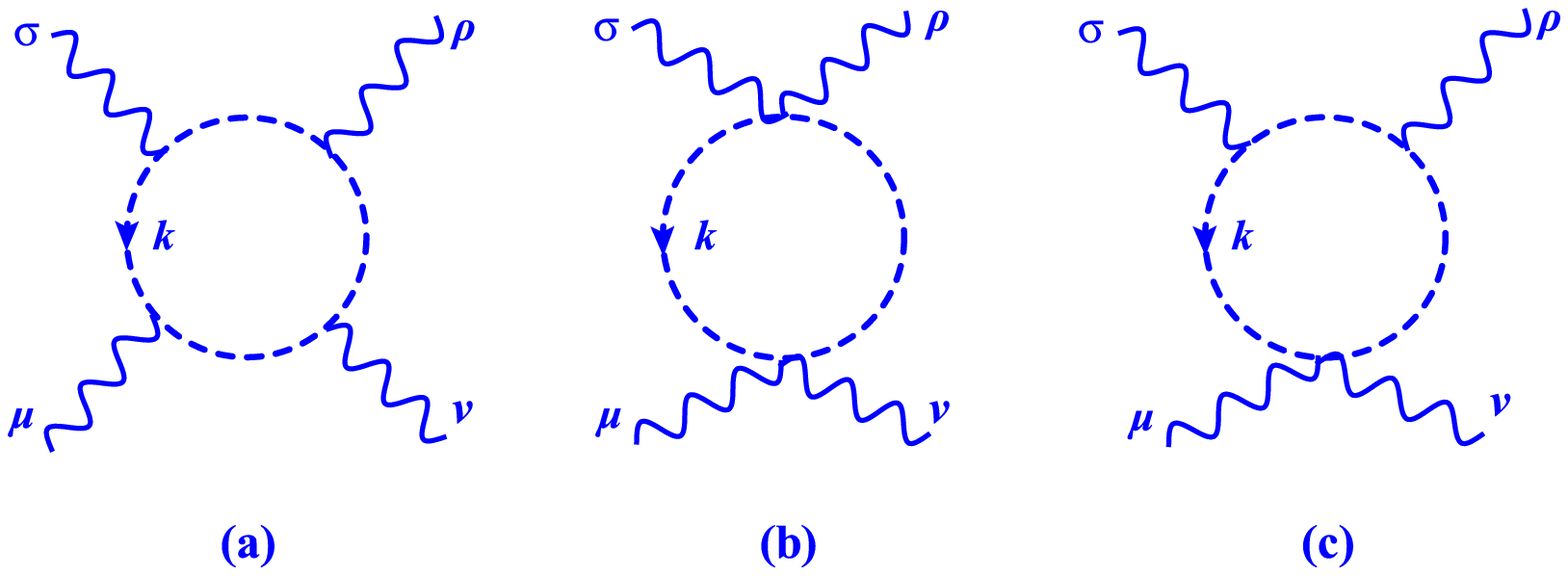}
 \centering\caption{Relevant graphs for the induced $AAAA$-term.}
\label{oneloop3}
\end{figure}

The full contribution to the $\left\langle AA AA \right\rangle$ part is determined by considering three different types of diagrams that are depicted in Fig.~\ref{oneloop3}. Since all of these graphs have 4 external bosonic legs, 24 different permutations for each graph must be considered in order to obtain the fully symmetrized contribution.
Hence, the full contribution can be formally written as
\begin{align}
 \Gamma^{\mu\nu\rho\sigma}_{\rm total}= \Gamma^{\mu\nu\rho\sigma}_{(a)} +
 \Gamma^{\mu\nu\rho\sigma}_{(b)} +\Gamma^{\mu\nu\rho\sigma}_{(c)} =\sum_{i=1}^{24}\Gamma^{\mu\nu\rho\sigma}_{(a,i)}+
  \sum_{i=1}^{24}\Gamma^{\mu\nu\rho\sigma}_{(b,i)}+
   \sum_{i=1}^{24}\Gamma^{\mu\nu\rho\sigma}_{(c,i)}.
    \label{eq:c17}
\end{align}
We shall present next the explicit discussion for the first contribution of each graph, whereas the remaining graphs are obtained by a direct permutation of momenta and spacetime indices.
%%%%%%%%%%%%%%%%%%%%%%%%%%%%%%%%%%%%%%%%%%%%%%%%%%%%%%%%%%%%%%%%%%%%%%%%%%%%%%%%%%%%%
%%%%%%%%%%%%%%%%%%%%%%%%%%%%%%%%%%%%%%%%%%%%%%%%%%%%%%%%%%%%%%%%%%%%%%%%%%%%

The box diagram contribution represented in graph (a) has the following expression
\begin{align}
\Pi^{\mu\nu\rho\sigma}_{(a,1)}&=e^{4}\int \frac{d^{d}k}{(2\pi)^{d}} \frac{(p+2k)^{\mu}(2p+q+2k)^{\nu}(2p+2q+s+2k)^{\rho}(p+q+s+2k)^{\sigma}}{[(p+q+s+k)^{2} -m^{2}][(p+q+k)^{2}-m^{2}][(p+k)^{2}-m^{2}][k^{2}-m^{2}]}
e^{\frac{i}{2}p \wedge q}e^{\frac{i}{2}(p+q)\wedge s},
 \label{eq:c13}
\end{align}
where we have labeled the momenta $(p,q,s,r)$ accordingly to the spacetime indices of the external legs $(\mu,\nu,\rho,\sigma)$.
Moreover we have adopted the notation, in order to satisfy the energy-momentum conservation, where the momenta flow satisfies the relation $r = p+q+s$.
The remaining contributions from the other 23 box diagrams, coming from the S-matrix expansion, can easily be obtained from the equation  \eqref{eq:c13} by considering the respective permutation.
Next, we have the contribution from the bubble diagram represented in (b), which is given by
\begin{align}
\Pi^{\mu\nu\rho\sigma}_{(b,1)} =e^{4} \int  \frac{d^{d}k}{(2\pi)^{d}} \frac{\eta^{\mu\nu}\eta^{\rho\sigma}}{[(p+q+k)^{2}-m^{2}][k^{2}-m^{2}]}\cos\big(\frac{p \wedge q}{2}\big)\cos\big(\frac{s\wedge (p+q)}{2}\big).
 \label{eq:c14}
\end{align}
At last, the triangle contribution shown in graph (c) is written as
\begin{align}
\Pi^{\mu\nu\rho\sigma}_{(c,1)} =-e^{4}\int \frac{d^{d}k}{(2\pi)^{d}} \frac{\eta^{\mu\nu}(2p+2q+s+2k)^{\rho}(p+q+s+2k)^{\sigma}}{[(p+q+s+k)^{2}-m^{2}][(p+q+k)^{2}-m^{2}][k^{2}-m^{2}]}\cos\big( \frac{p \wedge q}{2}\big) e^{\frac{i}{2} (p+q) \wedge s}.
 \label{eq:c15}
\end{align}
The Feynman expressions of the graphs (a), (b) and (c) show that all of them are planar, making the evaluation of the momentum integration easier by dimensional regularization.
Hence the resulting expressions from the contributions \eqref{eq:c13} to \eqref{eq:c15}, evaluated in the highly noncommutative limit, where $p^2,q^2,s^2 \ll m^{2}$, are written as follows
\begin{align}
\Gamma^{\mu\nu\rho\sigma}_{(a,1)}&=\frac{1}{4}\times \frac{ie^{4}}{12\pi m}\Big(\eta^{\mu\nu}\eta^{\rho\sigma}+\eta^{\mu\rho}\eta^{\nu\sigma}+\eta^{\mu\sigma}\eta^{\nu\rho}\Big)
~e^{\frac{i}{2}p\wedge q}e^{\frac{i}{2}r\wedge s},\nonumber\\
%%%%%%%
\Gamma^{\mu\nu\rho\sigma}_{(b,1)}&= \frac{1}{2}\times\frac{ie^{4}}{8\pi m}\eta^{\mu\nu}\eta^{\rho\sigma}
\cos\big(\frac{p\wedge q}{2}\big)\cos\big(\frac{r\wedge s}{2}\big),\nonumber\\
%%%
\Gamma^{\mu\nu\rho\sigma}_{(c,1)}&=1\times \frac{-ie^{4}}{8\pi m} ~\eta^{\mu\nu}\eta^{\rho\sigma}\cos\big(\frac{p\wedge q}{2}\big)\cos\big(\frac{r\wedge s}{2}\big),
 \label{eq:c16}
\end{align}
where the coefficients $\frac{1}{4}$, $\frac{1}{2}$ and $1$ are the symmetry factors for the graphs (a), (b) and (c) of Fig.~\ref{oneloop3}, respectively.
We then apply to the results \eqref{eq:c16} all the 24 permutations, necessary to evaluate \eqref{eq:c17}, yielding
\begin{align}
\Gamma^{\mu\nu\rho\sigma}_{(a)}&=\frac{ie^{4}}{6\pi m}\Big(\eta^{\mu\nu}\eta^{\rho\sigma}+\eta^{\mu\rho}\eta^{\nu\sigma}+\eta^{\mu\sigma}\eta^{\nu\rho}\Big)
\Big[\cos\big(12\big)\cos\big(34\big)+\cos\big(13\big)\cos\big(24\big)+\cos\big(14\big)\cos\big(23\big)\Big],
\nonumber\\
\Gamma^{\mu\nu\rho\sigma}_{(b)}&=\frac{ie^{4}}{2\pi m}
\Big[\eta^{\mu\nu}\eta^{\rho\sigma}\cos\big(12\big)\cos\big(34\big)+\eta^{\mu\rho}
\eta^{\nu\sigma}\cos\big(13\big)\cos\big(24\big)+
\eta^{\mu\sigma}\eta^{\nu\rho}\cos\big(14\big)\cos\big(23\big)\Big],
\nonumber\\
\Gamma^{\mu\nu\rho\sigma}_{(c)}&=- \frac{ie^{4}}{\pi m}
\Big[\eta^{\mu\nu}\eta^{\rho\sigma}\cos\big(12\big)\cos\big(34\big)+\eta^{\mu\rho}
\eta^{\nu\sigma}\cos\big(13\big)\cos\big(24\big)+
\eta^{\mu\sigma}\eta^{\nu\rho}\cos\big(14\big)\cos\big(23\big)\Big],
 \label{eq:c18}
\end{align}
where we have introduced, by simplicity of the upcoming analysis, the following notation for the NC momenta product: $\left(12 \right) \equiv \left(\frac{p\wedge q}{2}\right)$, $\left(13 \right) \equiv \left(\frac{p\wedge s}{2}\right)$, $\left(14 \right) \equiv \left(\frac{p\wedge r}{2}\right)$, $\left(23 \right) \equiv \left(\frac{q\wedge s}{2}\right)$, $\left(24 \right) \equiv \left(\frac{q\wedge r}{2}\right)$, and $ \left(34\right) \equiv  \left(\frac{s\wedge r}{2}\right)$.
Finally, we substitute the results \eqref{eq:c18} into the Eq.~\eqref{eq:c17} to obtain the total one-loop contribution expression corresponding to the photon 4-point function
\begin{align}
\Gamma^{\mu\nu\rho\sigma}_{\rm total}=\frac{ie^{4}}{\pi m}\Bigg\{
&\frac{1}{6}\Big(\eta^{\mu\nu}\eta^{\rho\sigma}+\eta^{\mu\rho}\eta^{\nu\sigma}+\eta^{\mu\sigma}\eta^{\nu\rho}\Big) \Big[\cos(12)\cos(34)+\cos(14)\cos(23)+\cos(13)\cos(24)\Big]\nonumber\\
-&\frac{1}{2} \Big[\eta^{\mu\nu}\eta^{\rho\sigma}\cos(12)
\cos(34)+\eta^{\mu\rho}\eta^{\nu\sigma}\cos(13)\cos(24)+\eta^{\mu\sigma}\eta^{\nu\rho}\cos(14)\cos(23)\Big]\Bigg\}.
 \label{eq:c19}
\end{align}
We can verify whether the quartic vertex \eqref{eq:c19} satisfy the Ward identity in scalar QED. First, we consider the commutative limit, i.e. $\theta\rightarrow 0$, so that the one-loop contribution \eqref{eq:c19} is reduced to
\begin{equation}
\lim_{\theta\rightarrow 0}\Gamma^{\mu\nu\rho\sigma}_{\rm total}=\frac{ie^{4}}{\pi m}
\Big(\eta^{\mu\nu}\eta^{\rho\sigma}+\eta^{\mu\rho}\eta^{\nu\sigma}+\eta^{\mu\sigma}\eta^{\nu\rho}\Big) \Big(\frac{1}{2}+\frac{1}{2}-1\Big)=0,
 \label{eq:c20}
\end{equation}
showing that the photon quartic self-coupling in the order ${\cal{O}}(m^{-1})$ is absent in the Abelian theory; however higher order contributions could be nonvanishing, corresponding to nonlinear Euler-Heisenberg-like terms.
Moreover, in the case of the violation of the Ward identity, a nonvanishing result for the contribution \eqref{eq:c20} would generate a four photon interaction term in the effective action of the type
\begin{align}
\lim_{\theta\rightarrow 0}{\cal{S}}_{\rm eff}[AAAA]\sim\int d^{3}x~\Big(\eta^{\mu\nu}\eta^{\rho\sigma}+\eta^{\mu\rho}\eta^{\nu\sigma}+\eta^{\mu\sigma}\eta^{\nu\rho}\Big)A_{\mu}(x) A_{\nu}(x) A_{\rho}(x)A_{\sigma}(x),
 \label{eq:c21}
\end{align}
that is not explicitly gauge invariant. Hence, with the result \eqref{eq:c20} we conclude that the gauge invariance is satisfied in our analysis of ${\cal{O}}(m^{-1})$ terms at the one-loop approximation.

On the other hand, in the noncommutative case, as it is well known, we expect to find the relevant Feynman rule corresponding to the 4-photon interaction term at this order.
To accomplish that, we start working separately each one of the tensor terms present in the function $\Gamma^{\mu\nu\rho\sigma}_{\rm total}$  Eq.~\eqref{eq:c19}.
We shall illustrate the analysis for the terms proportional to $\eta^{\mu\nu}\eta^{\rho\sigma}$, the remaining terms can be evaluated in the same fashion.
Hence, by picking the pieces that are proportional to $\eta^{\mu\nu}\eta^{\rho\sigma}$ in \eqref{eq:c19}, we have that
\begin{equation}
\mathcal{I}^{\mu\nu,\rho\sigma}\equiv
\frac{ie^{4}}{6\pi m}\eta^{\mu\nu}\eta^{\rho\sigma}\Big[-2\cos(12)\cos(34)+ \cos(14)\cos(23)+\cos(13)\cos(24)\Big]. \label{eq:c22}
\end{equation}
The major work here consist in simplifying the trigonometric part of this function by making use of the energy-momentum conservation  $r = p+q+s$, which in the new notation is renamed as $4=1+2+3$, together with the manipulation of some trigonometric identities, e.g. $\cos\alpha\cos\beta =\cos(\alpha+\beta)+\sin\alpha\sin\beta$.
After some laborious but straightforward calculation, we arrive at the desired expression
\begin{equation}
\mathcal{I}^{\mu\nu,\rho\sigma}=\frac{ie^{4}}{6\pi m}\eta^{\mu\nu}\eta^{\rho\sigma}\Big[\sin(14)\sin(23)+\sin(13)\sin(24)\Big].
 \label{eq:c24}
\end{equation}
Similarly, we can apply the same process in order to simplify the remaining terms, proportional to $\eta^{\mu\rho}\eta^{\nu\sigma}$ and $\eta^{\mu\sigma}\eta^{\nu\rho}$, so  that it yields to
\begin{align}
\mathcal{I}^{\mu\rho,\nu\sigma}&=\frac{ie^{4}}{6\pi m}\eta^{\mu\rho}\eta^{\nu\sigma}\Big[\sin(12)\sin(34)-\sin(14)\sin(23)\Big],
\nonumber\\
\mathcal{I}^{\mu\sigma,\nu\rho}&=- \frac{ie^{4}}{6\pi m}\eta^{\mu\sigma}\eta^{\nu\rho}\Big[\sin(12)\sin(34)+\sin(13)\sin(24)\Big].
 \label{eq:c25}
\end{align}

Hence, by considering the results from our manipulations, Eqs.~\eqref{eq:c24} and \eqref{eq:c25}, we can rewrite \eqref{eq:c19} in a convenient form as the following
\begin{align}
\Gamma^{\mu\nu\rho\sigma}_{\rm total}=\frac{ie^{4}}{6\pi m}\bigg[&\Big(\eta^{\mu\nu}\eta^{\rho\sigma}-\eta^{\mu\sigma}\eta^{\nu\rho}\Big)\sin\big(\frac{p\wedge s}{2}\big)\sin\big(\frac{q\wedge r}{2}\big)\nonumber\\
+&\Big(\eta^{\mu\nu}\eta^{\rho\sigma}-\eta^{\mu\rho}\eta^{\nu\sigma}\Big)\sin\big(\frac{p\wedge r}{2}\big)\sin\big(\frac{q\wedge s}{2}\big)\nonumber\\
+&\Big(\eta^{\mu\rho}\eta^{\nu\sigma}-\eta^{\mu\sigma}\eta^{\nu\rho}\Big)\sin\big(\frac{p\wedge q}{2}\big)
\sin\big(\frac{s\wedge r}{2}\big)\bigg],
 \label{eq:c27}
\end{align}
where we have reintroduced the notation in terms of the external momenta $p,q,s,r$.
As we can observe, the expression inside the bracket corresponds exactly to the Feynamn vertex of the 4-photon interaction within the noncommutative $U_{\star}(1)$ gauge theory.

Finally, we can gather the leading ${\cal{O}}(m^{-1})$ contributions from the one-loop order parts related to the two, three and four-point functions, Eqs.~\eqref{eq:c3}, \eqref{eq:c11} and \eqref{eq:c27}, respectively, so that we can write the complete expression of the NC Maxwell action as
\begin{equation}
i{\cal{S}}_{\rm eff}\bigg|_{{\cal{O}}(m^{-1})}=-\frac{ie^{2}}{96\pi m}\int~d^{3}x~F_{\mu\nu}\star F^{\mu\nu},
\label{eq:c28}
\end{equation}
in which the field strength tensor in the NC framework is defined as $F_{\mu\nu}=\partial_{\mu}A_{\nu}-\partial_{\nu}A_{\mu}+ie[A_{\mu},A_{\nu}]_{\star}$.
As we have previously discussed, this action is manifestly $U_{\star}(1)$  gauge invariant under the transformation $U=e^{ie\lambda}_{\star}$, where the field strength has the following transformation law $F_{\mu\nu}\rightarrow U\star F_{\mu\nu}\star U^{-1}$.

Regarding the higher-derivative corrections to the 4-photon vertex \eqref{eq:c27}, corresponding to the next to leading order ${\cal{O}}(m^{-3})$ terms, we arrive at a result involving a long expression which can be found in the Appendix \ref{sec-appA}.
This ${\cal{O}}(m^{-3})$ result can be seen as the 3D version of the Euler-Heisenberg Lagrangian.
We notice that in the commutative limit, the gauge invariant field strength is defined as $f_{\mu\nu}=\partial_{\mu}A_{\nu}-\partial_{\nu}A_{\mu}$, so that the commutative version of the effective action \eqref{eq:c28} only receives contribution from the one-loop $\langle AA\rangle$ part, the remaining contributions are vanishing.
Thus, the commutative one-loop effective action in the presence of the higher-derivative term, at the next to leading order, is described as
 \begin{equation}
 \lim\limits_{\theta\rightarrow 0}i{\cal{S}}_{\rm eff}\bigg|_{{\cal{O}}(m^{-3})}=
 -\frac{ie^{2}}{96\pi m}\int d^{3}x~f_{\mu\nu}f^{\mu\nu}+\frac{ie^{2}}{1920\pi m^{3}}\int d^{3}x~
f_{\mu\nu}\Box f^{\mu\nu}
\end{equation}

Some comments about the result \eqref{eq:c28} are now in place.
Regarding discrete symmetries, following the aforementioned discussion in Sec. \ref{sec2.1}, it is easy to show that the above one-loop effective action is also invariant under all of the discrete symmetries and therefore we have not faced any anomalous symmetry at this order.

Since the effective action \eqref{eq:c28}, arising from the 2, 3 and 4-point functions at the order ${\cal{O}}(m^{-1})$, is exactly gauge invariant, it is possible to conclude that no further ${\cal{O}}(m^{-1})$ terms are generated from higher-order graphs with $n>4$ external photon legs.
According to this reasoning, we can also discuss the gauge invariance of the higher-derivative terms generated by considering the next to leading order ${\cal{O}}(m^{-3})$ of our expansion.
Actually, it is possible to make use of a dimensional analysis, based on arguments of gauge invariance, to establish the perturbative generation of all possible gauge invariant higher-derivative terms in the one-loop effective action \cite{Bufalo:2014ooa}.
As an example, if we consider all of the ${\cal{O}}(m^{-3})$ contributions up to the diagrams with $n=6$ external photon legs, we can generate the following effective higher-derivative Lagrangian
\begin{equation}
\mathcal{L}_{\rm hd} = \frac{1}{6\mu^2} \nabla _\mu F^{\mu \nu }\star \nabla^\lambda F_{\lambda \nu }
+\frac{1}{6\mu^2} \nabla _\lambda F^{\mu \nu }\star \nabla^\lambda F_{\mu \nu }
- \frac{e}{18\mu^2} F^{\mu \nu }\star  F_{\nu \lambda}\star  F^{\lambda } _{\;\;\;  \mu},
\end{equation}
where $\nabla _\mu  = \partial_\mu +i e \left[A_\mu, \,\right]_{\star} $ is the covariant derivative in the adjoint representation and $\mu \sim m$ is the mass scale of the theory.
This expression can been seen as a noncommutative extension of the Alekseev-Arbuzov-Baikov effective Lagrangian \cite{Quevillon:2018mfl,Alekseev:1981fu}.

One last comment about the photon effective action is in regard of some of the nonlinear contributions.
It is well known that either fermionic or scalar electrodynamics generate nonlinear corrections of quantum character to the photon dynamics \cite{Quevillon:2018mfl}.
In the case of a $(2+1)$ spacetime the effective Euler-Heisenberg Lagrangian density has the  appearance of fractional powers of the field strength \cite{Redlich:1983dv}
\begin{equation}
\mathcal{L}_{\rm EH} \sim  \left( e  \sqrt{B^2-E^2}  \right)^{\frac{3}{2}}.
\end{equation}
Hence, it is reasonable to expect that the coordinates noncommutativity would also present corrections to this nonlinear coupling term.

%%%%%%%%%%%%%%%%%%%%%%%%% %%%%%%%%%%%%%%%%%%%%%%%%%
 \section{Comparison with NC-QED$_{3}$}
\label{sec5}

In this section, we shall present a comparative discussion of the 2,3 and 4-point functions in the case of fermionic and bosonic matter fields coupled to the photon.
It is notable that the presence of the trace of $\gamma$ matrices in fermionic QED$_{3}$ leads to two sectors characterized as: odd and even in regard to parity symmetry.
The former is related to the induced Chern-Simons (CS) terms, appearing with odd powers of the fermion mass $m_e$, while the latter sector contributes to the induced Maxwell (M) terms, with even powers of $m_e$.
Now in the scalar framework, odd parity terms are absent, and only parity preserving terms are present in the induced effective action, which can be understood as the main difference between these two matter fields.
These results can be briefly described in terms of a mass expansion as the following:

\begin{center}
\begin{tabular}{|c|c|c|c|c|c|c|}
\hline
{\color{blue}Order} & {\color{blue}Induced action} & {\color{blue}NC-QED$_{3}$}& {\color{blue}NC-scalar QED$_{3}$} \\
\hline
${\cal{O}}(m^{0})$ & ordinary~NC-CS & $\checkmark$ & $\times$ \\
\hline
~${\cal{O}}(m^{-1})$ & ordinary~NC-M~ & $\checkmark$ & $\checkmark$ \\
\hline
~${\cal{O}}(m^{-2})$& higher-derivative~NC-CS & $\checkmark$ & $\times$ \\
\hline
~${\cal{O}}(m^{-3})$& higher-derivative~NC-M~ & $\checkmark$ & $\checkmark$\\
\hline
\vdots& \vdots& \vdots& \vdots\\
\hline
~~${\cal{O}}(m^{-2\ell})$& higher-derivative~NC-CS & $\checkmark$ & $\times$\\
\hline
~~~~~${\cal{O}}(m^{-2\ell-1})$& higher-derivative~NC-M~ & $\checkmark$ & $\checkmark$\\
\hline
\end{tabular}
\end{center}
Here some important comments are in order.
In NC-QED$_{3}$, the structure of the kinetic part, coming from the $n=2$ photon external legs analysis, has the contribution of two types of terms:
a CS-type parity violating term $e^{2}A \partial\Box^{\ell} A$, at the order ${\cal{O}}(m^{-2\ell})$, and a M-type parity preserving term $e^{2}A\partial\partial \Box^{\ell} A$, at the order ${\cal{O}}(m^{-2\ell-1})$.
We can observe that the mass dimension of the CS and M-type terms is given by $3+2\ell$ and $4+2\ell$, respectively.
Thus, we can conclude that in order to have a gauge invariant CS-type action at the order ${\cal{O}}(m^{-2\ell})$, it is necessary to consider all contributions originating from the graphs with $n=2,3,\ldots, 3+2\ell$ photon legs.
On the other hand, to obtain a gauge invariant M-type action at the order ${\cal{O}}(m^{-2\ell-1})$, it is necessary to consider all contributions arising from the relevant graphs with $n=2,3,\ldots, 4+2\ell$ photon legs.

Furthermore, in $(2+1)$ dimensions, the number of degrees of freedom of the charged boson and Dirac fermion (in the 2 dimensional representation) is equal and hence it is easy to see that the numerical coefficient appearing in the 2-point function  \eqref{eq:c7} would be the same as in the fermionic QED$_{3}$.
Now for the $n=3$ graphs, in the case of NC-QED$_{3}$, there are only the contribution of two triangle graphs in the one-loop order.
These contributions are the same as in the NC-scalar QED$_{3}$ because the additional graph is identically zero.
Thus, it is easy to realize that the final result in the parity preserving sector for both cases is the same \cite{Bufalo:2014ooa}.
The last type of diagrams is for $n=4$ legs, that for the fermionic case we have the contribution of the box diagram only, i.e. type (a).
The expression for this diagram at the leading order of ${\cal{O}}(m^{-1})$ is given by
\begin{equation}
\Gamma^{\mu\nu\rho\sigma}_{(a,1)}\bigg|_{\tiny\mbox{NC-QED}}=-\frac{ie^{4}}{3\pi m}\Big(\eta^{\mu\sigma}\eta^{\nu\rho}-2\eta^{\mu\rho}\eta^{\nu\sigma}+ \eta^{\mu\nu}\eta^{\rho\sigma}\Big)e^{\frac{i}{2}p\wedge q}e^{\frac{i}{2}r\wedge s}.
\end{equation}
By considering all of the 24 permutations and some manipulations we obtain the following
\begin{align}
\Gamma^{\mu\nu\rho\sigma}_{\rm total}\bigg|_{\tiny\mbox{NC-QED}}=\frac{4ie^4}{3\pi m}\bigg[ \Big(&\eta^{\mu\nu}\eta^{\rho\sigma}-\eta^{\mu\rho}\eta^{\nu\sigma}\Big)\sin{\big(\frac{p\wedge r}{2}\big)}\sin{\big(\frac{q\wedge s}{2}\Big)}\nonumber\\
+\Big(&\eta^{\mu\nu}\eta^{\rho\sigma}-\eta^{\mu\sigma}\eta^{\nu\rho}\Big)\sin{\big(\frac{q\wedge r}{2}\big)}\sin{\big(\frac{p\wedge s}{2}\big)}\nonumber\\
+\Big(&\eta^{\mu\rho}\eta^{\nu\sigma}-\eta^{\mu\sigma}\eta^{\nu\rho}\Big)\sin{\big(\frac{p\wedge q}{2}\big)}\sin{\big(\frac{s\wedge r}{2}\big)}\bigg],
\end{align}
which has the same tensor and momenta structure as the standard 4-photon vertex \eqref{eq:c27}, but with a different numerical coefficient.
In this case, we can conclude that the commutative limit of this function is also satisfied, and have a vanishing result as we would expect.
%%%%%%%%%%%%%%%%%%%%%%%%%%%%%%%%%%%%%%%%%%%%%%%%%%%%%%%%%%%%%%%%%%%%%%%
\section{Final remarks}
\label{conc}

In this paper we have considered the perturbative evaluation of the effective action for photon in the context of scalar QED in the $(2+1)$ noncommutative spacetime.
Our main interest was to determine to what extent the spin of the matter fields can change the effective action in the presence of the spacetime noncommutativity.
Since the number of degrees of freedom of the charged boson and Dirac fermion (in the 2 dimensional representation) is equal, the main difference between these fields is solely to the well known presence of different couplings in the case of scalar QED, implying in new types of graphs.
An important drawback from the scalar QED in terms of the induced effective action for the photon is the absence of parity violating terms, showing that no Chern-Simons terms are generated when charged scalar fields are considered.

The perturbative analysis followed the computation of the $\left\langle AA  \right\rangle$, $\left\langle AA A \right\rangle$ and $\left\langle AA AA \right\rangle$ vertex functions.
A more detailed and careful analysis was necessary to the computation of the 4 point vertex, where the process of symmetrization is rather intricate.
The evaluation of the noncommutative Maxwell action $\int F_{\mu\nu}\star F^{\mu\nu} $ was done by considering the highly noncommutative limit of these $1PI$ functions at order ${\cal{O}}(m^{-1})$.

In addition, we have considered the generation of ${\cal{O}}(m^{-3})$ terms, which are higher-derivative terms for the photon fields, and can be seen as the noncommutative generalization of the phenomenological Alekseev-Arbuzov-Baikov effective Lagrangian.
Another possible terms to be present in the gauge invariant photon's effective action are those nonlinear couplings, e.g. analogous to the Euler-Heisenberg action, thus within our study of NC-scalar QED, we would find noncommutative corrections to the fractional powers of the field strength that appear in the $(2+1)$ dimensional Euler-Heisenberg action.

%%%%%%%%%%%%%%%%%%%%%%%%%%%%%%%%%%%%%%%%% %%%%%%%%%%%%%%%%%%%%%%%%%%%%%%
%%%%%%%%%%%%%%%%%%%%%%%%%%%%%%%%%%%%%%%%%%%%%%%%%%%%%%%%%%%%%%%%%%%%%%%
 \subsection*{Acknowledgements}

 We are grateful to M.M. Sheikh-Jabbari for fruitful discussions and comments on the manuscript. R.B. acknowledges partial support  from Conselho Nacional de Desenvolvimento Cient\'ifico e Tecnol\'ogico (CNPq Projects No. 304241/2016-4 and 421886/2018-8 ) and Funda\c{c}\~{a}o de Amparo \`a Pesquisa do Estado de Minas Gerais (FAPEMIG Project No. APQ-01142-17).

%%%%%%%%%%%%%%%%%%%%%%%%%%%%%%%%%%%%%%%%%%%%%%%%%%%%%%%%%%%%%%%%%%%%%%%%%%%%%%%%%%%%%%
\appendix
\section{Higher-derivative corrections to the 4-photon vertex}
\label{sec-appA}
The higher-derivative contributions of the graphs, depicted in Fig.~\ref{oneloop3}, to the 4-photon vertex at the next to leading order, ${\cal{O}}(m^{-3})$, are described as the following:
\begin{eqnarray}
\Gamma^{\mu\nu\rho\sigma}_{\tiny\mbox{hd}}=\Gamma^{\mu\nu\rho\sigma}_{\tiny\mbox{hd}~(a)}
+\Gamma^{\mu\nu\rho\sigma}_{\tiny\mbox{hd}~(b)}
+\Gamma^{\mu\nu\rho\sigma}_{\tiny\mbox{hd}~(c)}
\end{eqnarray}
For simplicity, we define $\Gamma^{\mu\nu\rho\sigma}_{\tiny\mbox{hd}~(a)}=\frac{ie^4}{240 \pi m^3 }\widetilde{\Gamma}^{\mu\nu\rho\sigma}_{\tiny\mbox{hd}~(a)}$ and $\Gamma^{\mu\nu\rho\sigma}_{\tiny\mbox{hd}~(b+c)}=-\frac{ie^{4}}{24\pi m^{3}}\widetilde{\Gamma}^{\mu\nu\rho\sigma}_{\tiny\mbox{hd}~(b+c)}$ in which
\begin{align*}
\widetilde{\Gamma}^{\mu\nu\rho\sigma}_{\tiny\mbox{hd}~(a)}=\Bigg\{&\eta^{\mu \nu} \Big[ - 2 p^{\sigma} p^{\rho} + 4 q^{\sigma} p^{\rho} -
6 s^{\sigma} p^{\rho} - q^{\rho} p^{\sigma} - s^{\rho} p^{\sigma} + 2
q^{\sigma} s^{\rho} + q^{\rho} q^{\sigma} - 3 q^{\rho} s^{\sigma} - 4
s^{\rho} s^{\sigma}\Big]
\nonumber\\
 +& \eta^{\mu \rho} \Big[ - 2 \left(p^{\sigma} + 3
q^{\sigma} - 2 s^{\sigma} \right) p^{\nu} - q^{\nu} p^{\sigma} -
s^{\nu} p^{\sigma} - 4 q^{\nu} q^{\sigma} - 3 s^{\nu} q^{\sigma} + 2
q^{\nu} s^{\sigma} + s^{\nu} s^{\sigma}\Big]
\nonumber\\
+& \eta^{\mu \sigma} \Big[4 \left(2
p^{\rho} + q^{\rho} + s^{\rho} \right) p^{\nu} + 4 q^{\nu} p^{\rho} +
4 s^{\nu} p^{\rho} + 2 q^{\rho} s^{\nu} + q^{\nu} q^{\rho} + 2 q^{\nu}
s^{\rho} + s^{\nu} s^{\rho}\Big]
\nonumber\\
+& \eta^{\nu \rho} \Big[ - 2 p^{\sigma} p^{\mu} -
q^{\sigma} p^{\mu} - s^{\sigma} p^{\mu} - q^{\mu} p^{\sigma} - s^{\mu}
p^{\sigma} + q^{\mu} q^{\sigma} - 3 s^{\mu} q^{\sigma} + s^{\mu}
s^{\sigma} - 3 q^{\mu} s^{\sigma}\Big]
\nonumber\\
+& \eta^{\nu \sigma} \Big[ - 2 p^{\rho}
p^{\mu} - q^{\rho} p^{\mu} - s^{\rho} p^{\mu} + 4 s^{\mu} p^{\rho} - 6
q^{\mu} p^{\rho} + 2 q^{\rho} s^{\mu} - 4 q^{\mu} q^{\rho} + s^{\mu}
s^{\rho} - 3 q^{\mu} s^{\rho}\Big]
\nonumber\\
 +& \eta^{\rho \sigma} \Big[ - 2 p^{\nu} p^{\mu}
- q^{\nu} p^{\mu} - s^{\nu} p^{\mu} + 4 q^{\mu} p^{\nu} - 6 s^{\mu}
p^{\nu} + q^{\mu} q^{\nu} + 2 q^{\mu} s^{\nu} - 3 s^{\mu} q^{\nu} - 4
s^{\mu} s^{\nu}\Big]
\nonumber\\
+&
\Big(\eta^{\mu\nu}\eta^{\rho\sigma}+\eta^{\mu\rho}\eta^{\nu\sigma}+\eta^{\mu\sigma}\eta^{\nu\rho}\Big)
\Big[4p^{2}+3q^{2}+3s^{2}+4(p.q)+4(p.s)+2(q.s)\Big]\Bigg\}\cos (\frac{r\wedge s+q\wedge p}{2})\nonumber\\
\end{align*}
%%%%%%%
\begin{align}
+\Bigg\{&\eta^{\mu \nu}
\Big[p^{\rho} p^{\sigma} - p^{\rho} q^{\sigma} - 3 p^{\rho} s^{\sigma} +
4 p^{\sigma} q^{\rho} + 2 p^{\sigma} s^{\rho} - 2 q^{\rho} q^{\sigma}
- q^{\sigma} s^{\rho} - 6 q^{\rho} s^{\sigma} - 4 s^{\rho} s^{\sigma}\Big]
\nonumber\\
+&\eta^{\mu \rho}\Big[(p^{\sigma} - q^{\sigma} - 3  s^{\sigma})
p^{\nu} - p^{\sigma} q^{\nu} - 3 p^{\sigma} s^{\nu} - 2 q^{\nu}
q^{\sigma} - q^{\sigma} s^{\nu} - q^{\nu} s^{\sigma} + s^{\nu}
s^{\sigma}\Big]
\nonumber\\
+&\eta^{\mu \sigma} \Big[- (4 p^{\rho} + 6 q^{\rho} + 3
 s^{\rho}) p^{\nu} - p^{\rho} q^{\nu} + 2 p^{\rho} s^{\nu} + 4
q^{\rho} s^{\nu} - 2 q^{\nu} q^{\rho} - q^{\nu} s^{\rho} + s^{\nu}
s^{\rho})
\nonumber\\
+&\eta^{\nu \rho}\Big[ - 4 p^{\mu} p^{\sigma} - p^{\mu}
q^{\sigma} + 2 p^{\mu} s^{\sigma} - 6 p^{\sigma} q^{\mu} - 3
p^{\sigma} s^{\mu} - 2 q^{\mu} q^{\sigma} - q^{\sigma} s^{\mu} + 4
q^{\mu} s^{\sigma} + s^{\mu} s^{\sigma}\Big]
\nonumber\\
 +&\eta^{\nu \sigma}\Big[p^{\mu}
p^{\rho} + 4 p^{\mu} q^{\rho} + 2 p^{\mu} s^{\rho} + 4 p^{\rho}
q^{\mu} + 2 p^{\rho} s^{\mu} + 4 q^{\rho} s^{\mu} + 8 q^{\mu} q^{\rho}
+ 4 q^{\mu} s^{\rho} + s^{\mu} s^{\rho}\Big]
\nonumber\\
+&\eta^{\rho \sigma}\Big[p^{\mu}
p^{\nu} - p^{\mu} q^{\nu} - 3 p^{\mu} s^{\nu} + 4 p^{\nu} q^{\mu} + 2
p^{\nu} s^{\mu} - 2 q^{\mu} q^{\nu} - q^{\nu} s^{\mu} - 6 q^{\mu}
s^{\nu} - 4 s^{\mu} s^{\nu}\Big]
\nonumber\\
+ &\Big(\eta^{\mu\nu}\eta^{\rho\sigma}+\eta^{\mu\rho}\eta^{\nu\sigma}+\eta^{\mu\sigma}\eta^{\nu\rho}\Big)
\Big[3 p^{2} + 4 q^{2} + 3 s^{2} + 4(p.q) + 2(p.s) + 4(q.s)\Big] \Bigg\}\cos (\frac{p\wedge r+q\wedge s}{2})\nonumber\\
+\Bigg\{&
\eta^{\mu \nu}\Big[p^{\rho} p^{\sigma} - 3 p^{\rho} q^{\sigma} - p^{\rho} s^{\sigma} -
3 p^{\sigma} q^{\rho} - p^{\sigma} s^{\rho} + q^{\rho} q^{\sigma} -
q^{\sigma} s^{\rho} - q^{\rho} s^{\sigma} - 2 s^{\rho} s^{\sigma}\Big]
\nonumber\\
+& \eta^{\mu \rho}\Big[(p^{\sigma} - 3 q^{\sigma} -  s^{\sigma})
p^{\nu} + 2 p^{\sigma} q^{\nu} + 4 p^{\sigma} s^{\nu} - 4 q^{\nu}
q^{\sigma} - 6 q^{\sigma} s^{\nu} - q^{\nu} s^{\sigma} - 2 s^{\nu}
s^{\sigma}\Big]
\nonumber\\
 +& \eta^{\mu \sigma}\Big[ -(4 p^{\rho} - 2 q^{\rho} +
s^{\rho}) p^{\nu} - 3 p^{\rho} q^{\nu} - 6 p^{\rho} s^{\nu} + 4
q^{\rho} s^{\nu} + q^{\nu} q^{\rho} - q^{\nu} s^{\rho} - 2 s^{\nu}
s^{\rho}\Big]
\nonumber\\
+& \eta^{\nu \rho}\Big[ - 4 p^{\mu} p^{\sigma} + 2 p^{\mu}
q^{\sigma} - p^{\mu} s^{\sigma} - 3 p^{\sigma} q^{\mu} - 6 p^{\sigma}
s^{\mu} + 4 q^{\sigma} s^{\mu} + q^{\mu} q^{\sigma} - q^{\mu}
s^{\sigma} - 2 s^{\mu} s^{\sigma}\Big]
\nonumber\\
+&\eta^{\nu \sigma}\Big[p^{\mu} p^{\rho} -
3 p^{\mu} q^{\rho} - p^{\mu} s^{\rho} + 2 p^{\rho} q^{\mu} + 4
p^{\rho} s^{\mu} - 4 q^{\mu} q^{\rho} - 6 q^{\rho} s^{\mu} - q^{\mu}
s^{\rho} - 2 s^{\mu} s^{\rho}\Big]
\nonumber\\
+& \eta^{\rho \sigma}\Big[p^{\mu} p^{\nu} + 2
p^{\mu} q^{\nu} + 4 p^{\mu} s^{\nu} + 2 p^{\nu} q^{\mu} + 4 p^{\nu}
s^{\mu} + 4 q^{\nu} s^{\mu} + q^{\mu} q^{\nu} + 4 q^{\mu} s^{\nu} + 8
s^{\mu} s^{\nu}\Big]
\nonumber\\
 +& \Big(\eta^{\mu\nu}\eta^{\rho\sigma}+\eta^{\mu\rho}\eta^{\nu\sigma}+\eta^{\mu\sigma}\eta^{\nu\rho}\Big)
 \Big[3p^{2} + 3 q^{2} + 4 s^{2} + 2(p.q)+4(p.s)+4(q.s)\Big]\Bigg\}\cos(\frac{p\wedge s+r\wedge q}{2})\nonumber\\
\end{align}
\begin{align}
\widetilde{\Gamma}^{\mu\nu\rho\sigma}_{\tiny\mbox{hd}~(b+c)}=
\Bigg\{&\Big[\big(p^2+q^2+p.q+s.r\big)\eta^{\mu\nu}\eta^{\rho\sigma}+p^{\nu}q^{\mu}\eta^{\rho\sigma}
-r^{\rho}s^{\sigma}\eta^{\mu\nu}\Big]\cos(\frac{p\wedge q}{2})\cos(\frac{s\wedge r}{2})
\nonumber\\
+&
\Big[\big(p^2+s^2+p.s+q.r\big)\eta^{\mu\rho}\eta^{\nu\sigma}+p^{\rho}s^{\mu}\eta^{\nu\sigma}
-r^{\nu}q^{\sigma}\eta^{\mu\rho}\Big]\cos(\frac{p\wedge s}{2})\cos(\frac{q\wedge r}{2})
\nonumber\\
+&
\Big[\big(q^2+s^2+q.s+p.r\big)\eta^{\mu\sigma}\eta^{\nu\rho}+q^{\rho}s^{\nu}\eta^{\mu\sigma}
-r^{\mu}p^{\sigma}\eta^{\nu\rho}\Big]\cos(\frac{p\wedge r}{2})\cos(\frac{q\wedge s}{2})
\Bigg\}
\end{align}

\newpage
%%%%%%%%%%%%%%%%%%%%%%%%%%%%%%%%%%%%%%%%%%%%%%%%%%%%%%%%%%%%%%%%%%%%%%%%%%%%%%%%%%%%%%%%%%%%%%%%%%%%%%%%%%%%%%%%%%%%%%%%%%%%%%%%%%%%%%%%%%%%%%%%%%%%%%%%%%%%%%%%%%%%%%%%%%%%%%%%%%%%%%%%%%%%%%%%%%%%%%%%%%%%%%%%%%%%%%

\global\long\def\link#1#2{\href{http://eudml.org/#1}{#2}}
 \global\long\def\doi#1#2{\href{http://dx.doi.org/#1}{#2}}
 \global\long\def\arXiv#1#2{\href{http://arxiv.org/abs/#1}{arXiv:#1 [#2]}}
 \global\long\def\arXivOld#1{\href{http://arxiv.org/abs/#1}{arXiv:#1}}

%%%%%%%%%%%%%%%%%%%%%%%%%%%%%%%%%%%%%%%%%%%%%%%%%%%%%%%%%%%%%%%%%%%%%%%%%%%%%%%%%%%%%%%%%%%%%%%%%%%%%%%%%%%

\end{document}